\documentclass{iau}
\usepackage{graphicx} 
\usepackage{hyperref}

\newcommand{\xmm}{{\em XMM-Newton}}
\newcommand{\cxo}{{\em Chandra}}

\def \etal   {\hbox{et~al.\/}}

\title[IAUS291.~~SXP 1062: a slow X-ray pulsar] 
{The slow X-ray pulsar SXP\,1062 and associated supernova 
remnant in the Wing of the Small Magellanic Cloud } 

\author[L.~M.~Oskinova \etal\ ]  
{L. M. Oskinova$^1$, M. A. Guerrero$^2$, V. H\'enault-Brunet$^3$, W. Sun$^4$, 
Y.-H. Chu$^5$, C. Evans$^3$, J. S. Gallagher, III$^6$,
R. A. Gruendl$^5$ and \\J. Reyes-Iturbide$^7$}
\affiliation{$^1$Institute for Physics and Astronomy, University of Potsdam, 14476 Potsdam, Germany,
\\ email: {\tt lida@astro.physik.uni-potsdam.de} \\[\affilskip]
$^2$Instituto de Astrofisica de Andalucia, IAA-CSIC,  Granada, Spain\\[\affilskip]
$^3$Scottish Universities Physics Alliance (SUPA), Institute for Astronomy, 
University of Edinburgh, UK\\[\affilskip]
$^4$Department of Astronomy, Nanjing University, Nanjing, 210093 Jiangsu, China\\[\affilskip]
$^5$Department of Astronomy, University of Illinois, 1002 West Green St., Urbana, IL 61801, USA\\[\affilskip]
$^6$Department of Astronomy, University of Wisconsin-Madison, Madison, WI 53706, USA\\[\affilskip]
$^7$Escuela Superior de Fisica y Matematicas, IPN, U.P. Adolfo L{\'o}pez Mateos, C.P. 07738 D.F., Mexico}

\pubyear{2012}
\volume{291}  
\jname{\mbox{Neutron Stars and Pulsars: Challenges and Opportunities after 80 years}}
\editors{J. van Leeuwen, ed.} 
\begin{document}

\maketitle

\begin{abstract}
SXP\,1062 is an exceptional case of a young neutron star in a wind-fed 
high-mass X-ray binary associated with a supernova remnant. A unique combination 
of measured spin period, its derivative, luminosity and young age makes 
this source a key probe for the physics of accretion and neutron star evolution.
Theoretical models proposed to explain the properties of SXP\,1062 shall be
tested with new data.
\end{abstract}


\firstsection 
\section{Introduction}
Neutron stars are the end products of  massive-star's evolution. Stars 
with an initial mass in excess of $\sim 8\,M_\odot$ end their life in a
core-collapse supernova (SN) explosion giving birth to a degenerate
compact object -- a neutron star (NS) or a black hole.  While the majority 
of massive stars are born in binary systems, only a small fraction ($\sim 10$\%)
of binaries survive the SN explosion, leaving a normal star and a
compact  object in a binary (e.g.\ Iben \& Tutukov 1996; Popov \&
Prokhorov 2006). At some point in binary evolution, the
compact object will  accrete matter from its companion entering the
high-mass X-ray binary (HMXB) stage. The detection of pulsations  from
an accreting X-ray source provides strong evidence that the  compact
object is a NS. The majority of HMXBs consist of a NS and a Be-type
star; these objects are called  BeXB (see recent review by Reig 2011). 
The Be star wind and disk feed the  NS making it an X-ray pulsar.

\begin{figure}[t]
\begin{center}
 \includegraphics[width=3.2in]{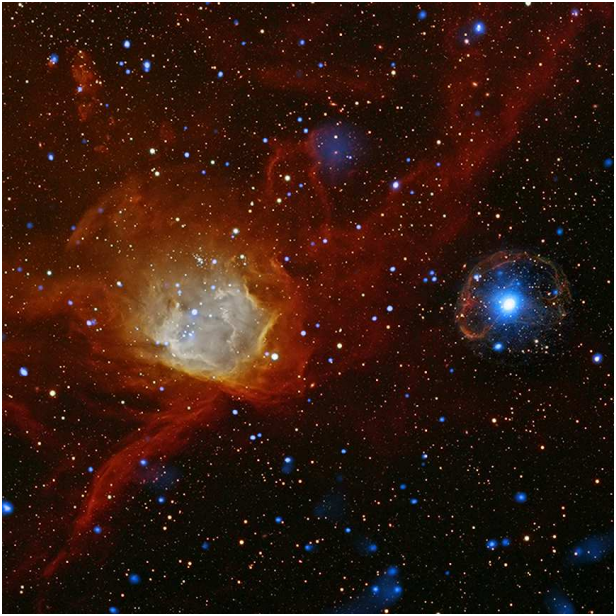} 
  \caption{Combined X-ray (blue) and optical (red, yellow) image of NGC\,602 and the SNR 
around the X-ray pulsar SXP\,1062. North is up, east is to the left. Image size is 14$'$, 
see \url{http://chandra.harvard.edu/photo/2011/sxp1062/}} 
    \label{fig1}
 \end{center}
\end{figure}

Pulsar spin  periods in the range 1-1000\,s can be explained according to 
the current  understanding of the NS  spin evolution (Reig 2011). 
The spin evolution can be divided into three key phases, characterized by a 
different energy release mechanism. These phases are known as the pulsar  phase, 
the propeller phase, and the accretor phase.  The spin period  $P_{\rm eq}$ is 
reached when the centrifugal and gravity forces balance. In principle, 
$P_{\rm eq}$ is the maximum spin period for a given mass-accretion rate. 
The mass-loss rates from OB-type stars are reasonably well known, despite 
the fact that the stellar winds have inhomogeneous structures 
(Oskinova \etal\ 2007, \v{S}urlan \etal\ 2012) 

In this canonical model, long periods in excess of 1000\,s can be
achieved if  $B>10^{14}$\,G or
$\dot{M}<10^{12}$\,g\,s$^{-1}$. On the other hand, a new model for 
wind accretion (Shakura \etal\ 2012) allows 
long spin periods and high period derivatives even for standard magnetic
fields. The model predicts specific correlations between the behavior 
of the spin and luminosity. An alternative model to explain long-period 
pulsars with $P_{\rm spin}\gg P_{\rm eq}$ comes from Ikhsanov (2007), who
postulates that prior to the accretion powered phase, a subsonic
propeller phase may take place. Steady accretion under the condition
$P_{\rm spin}>P_{\rm eq}$ can be  realized when the cooling of the
envelope plasma dominates the energy input. 

\section{SXP\,1062}
Until  recently no source was known that allows direct study of
various aspects  of the theory. 
The  situation changed dramatically with the discovery
of the BeXB SXP\,1062  (H\'enault-Brunet \etal\ 2012). It was discovered 
during \xmm\  and \cxo\ observations  in April-March 2010. 
SXP\,1062 is the first HMXB pulsar firmly associated with a supernova
remnant (SNR, see Fig.\,1). The SNR SXP\,1062 was discovered in  
H$\alpha$ and [O{\sc iii}] filter images (H\'enault-Brunet \etal\ 2012). 
The \cxo\ and \xmm\ X-ray images show that the SNR is filled with X-rays. 

The key observational parameters of SXP\,1062 were obtained 
(X-ray pulse period  $P= 1062$\,s and X-ray spectrum) using both
\cxo\ and \xmm\ observations. Based on the \xmm\ observations, Haberl \etal\ 
(2012) established the  spin-down rate of SXP\,1062  as 
$\dot{P}\approx 100$\,s\,yr$^{-1}$. The association of the pulsar with 
a SNR constrains the age of the accreting NS to 10-40\,kyr. 
Such young long period pulsars are theoretically not expected. 

\begin{figure}[t]
\begin{center}
 \includegraphics[width=3.2in]{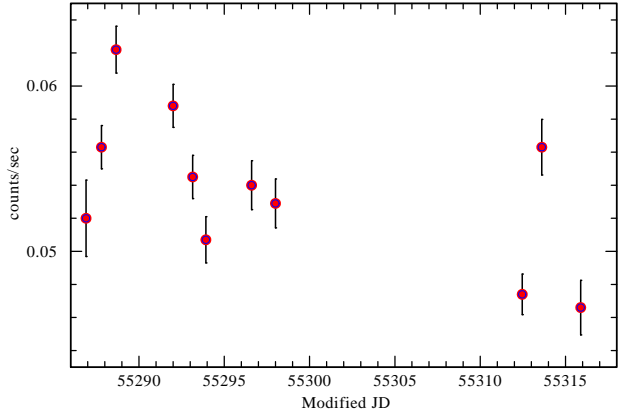} 
  \caption{The Chandra ACIS-I light-curve of SXP\,1062 in 0.4-10.0\,keV
band} 
    \label{fig2}
 \end{center}
\end{figure}

Figure\,1 shows the location of SXP\,1062 relative to the young star forming region 
NGC\,602. Some of the most massive stars in the SMC are identified in this cluster
(Evans \etal\ 2012).
The spectral type of the optical companion was confirmed spectroscopically as 
B0IIIe  star. Interestingly, some ``normal'' B-type stars have strong magnetic fields 
(Oskinova \etal\ 2011, and ref.\ therein), therefore an accretion of a magnetized stellar wind 
cannot be excluded. There are indications that SXP\,1062 has long term 
X-ray variability. Figure\,2 shows the X-ray light curve of SXP\,1062 based on 
11 \cxo\ exposures obtained in 2010.

Presently, four different explanations for the nature of SXP\,1062 have been proposed.
(1) Haberl \etal\ (2012) suggested that the NS in SXP\,1062 might have been born 
rotating unusually slowly. If this is the case, no field decay is required.
(2)  Popov \& Turola (2012) suggested that the NS was born as a
 magnetar, with an initial magnetic field $B>10^{14}$\,G. The strong magnetic braking  
and field dissipation led to the low $P$ and high $\dot{P}$ as observed.  
(3) Ikhsanov (2012) proposed that the accretion of magnetized matter can
lead to the observed low $P$, while the initial and the current magnetic 
field strength does not exceed $6\times 10^{13}$\,G. (4) Fu \& Li (2012) suggested 
that SXP\,1062 may be an accreting magnetar, with a present-day field $B>10^{14}$\,G.
New observations are needed to discriminate among these models and establish 
the true nature of SXP\,1062.

\end{document}